\documentclass[twocolumn,english,aps,prb,floats]{revtex4-2}
\usepackage[T1]{fontenc}
\usepackage[latin9]{inputenc}
\usepackage{babel}
\usepackage{amsmath}
\usepackage{amssymb}
\usepackage{graphicx}
\usepackage{esint}
\usepackage[unicode=true,
 bookmarks=false,
 breaklinks=false,pdfborder={0 0 1},backref=section,colorlinks=false]
 {hyperref}

\makeatletter

\newcommand{\lyxdot}{.}

\usepackage{babel}

\@ifundefined{textcolor}{}{%
 \definecolor{BLACK}{gray}{0}
 \definecolor{WHITE}{gray}{1}
 \definecolor{RED}{rgb}{1,0,0}
 \definecolor{GREEN}{rgb}{0,1,0}
 \definecolor{BLUE}{rgb}{0,0,1}
 \definecolor{CYAN}{cmyk}{1,0,0,0}
 \definecolor{MAGENTA}{cmyk}{0,1,0,0}
 \definecolor{YELLOW}{cmyk}{0,0,1,0}
}

\usepackage{ifpdf}\usepackage{bm}

\makeatother

\begin{document}
\title{Solid-Liquid Transition in a Skyrmion Matter}
\author{Dmitry A. Garanin and Eugene M. Chudnovsky}
\affiliation{Physics Department, Herbert H. Lehman College and Graduate School,
The City University of New York, 250 Bedford Park Boulevard West,
Bronx, New York 10468-1589, USA }
\date{\today}
\begin{abstract}
We report Monte-Carlo studies of the orientational order and melting
of a 2D skyrmion lattice containing more than one million spins. Two
models have been investigated, a microscopic model of lattice spins
with Dzyaloshinskii-Moryia interaction that possesses skyrmions, and
the model in which skyrmions are treated as point particles with repulsive
interaction derived from a spin model. They produce similar results.
The skyrmion lattice exhibits a sharp one-step transition between
solid and liquid phases on temperature and the magnetic field. This
solid-liquid transition is characterized by the kink in the magnetization.
The field-temperature phase diagram is computed. We show that the
application of the field gradient to a 2D system of skyrmions produces
a solid-liquid interface that must be possible to observe in experiments. 
\end{abstract}
\maketitle

\section{Introduction}

Skyrmions \cite{belpol75jetpl} have been among the most studied objects
in magnetic systems in recent years due to their potential for developing
topologically protected nanoscale information carriers \citep{Nagaosa2013,Zhang2015,Klaui2016,Hoffmann-PhysRep2017,Fert-Nature2017}.
In materials lacking inversion symmetry, individual skyrmions are
usually stabilized by Dzyaloshinskii-Moriya interaction (DMI) \citep{Bogdanov1989,Bogdanov94,Bogdanov-Nature2006,Heinze-Nature2011,Boulle-NatNano2016,Leonov-NJP2016}.
Frustrated exchange interactions \citep{Leonov-NatCom2015,Zhang-NatCom2017},
magnetic anisotropy \citep{IvanovPRB06,Lin-PRB2016}, disorder \citep{CG-NJP2018},
and geometrical confinement \citep{Moutafis-PRB2009} provide additional
mechanisms for the stabilization of skyrmions.

Bogdanov and Hubert \citep{Bogdanov94} pioneered studies of two-dimensional
(2D) lattices of magnetic vortices in systems with DMI. They obtained
the magnetic phase diagram separating periodic arrangements of vortices
from laminar domains. Skyrmion lattices were initially observed by
the real-space Lorentz transmission electron microscopy \citep{Yu2010}
in FeCoSi films. Many studies that followed confirmed reported transitions
between uniformly magnetized states, laminar domains, and skyrmion
crystals on temperature and the magnetic field, see, e.g., Refs. \citep{GCZZ-MMM2021,Dohi-ARCMP2022},
and references therein.

More recently, it was realized that skyrmion lattices provide a new
area for the studies of 2D melting \citep{Nishikawa-PRB2019,Huang-Nat2020,Balaz-PRB2021,GC-PRB2023}.
This general problem of statistical physics emerged in the 1970s with
the seminal works of Kosterlitz and Thouless \citep{KT}, Halperin
and Nelson \citep{HN-PRL1978,NH-PRB1979}, and Young \citep{Young-PRB1979}
(see Ref. \citep{Strandburg} for an early review). It was proposed
that, in line with the Kosterlitz-Thouless theory of 2D melting, a
2D solid melts via unbinding of defects but in a two-step manner via
the so-called KTHNY scenario named after the above authors. In that
scenario, a 2D solid first undergoes a transition into an intermediate
hexatic phase via unbinding of dislocations. It is characterized by
the exponential decay of translational correlations and the algebraic
decay of orientational correlations. Then, on further warming, it
melts into a true liquid state with exponential decay of all correlations
via unbinding of disclinations.

These theoretical predictions were confirmed by several numerical
studies \citep{DC-PRB1995,Li-PRE2019} and in experiments on colloidal
particles \citep{Murray-PRB1990,Keim}, see Ref.\ \citep{Kosterlitz-Review2016}
for review. Nevertheless, the phase diagram of the 2D melting is far
from being settled. Early studies that used molecular dynamics \citep{Broughton1982}
emphasized the importance of the competition between long-wave fluctuations,
which are particularly important in 2D, and short-wave phonons that
can drive a conventional first-order melting in 2D as they do in 3D.
Subsequently, it was established that the melting scenario was not
universal but depended on the interaction potential \citep{Broughton1982,Tsiok2022,Kapfer2015}
and the symmetry of the lattice \citep{Janke1988,Dietel2006}.

The debate related to the melting scenario has proliferated into recent
studies of skyrmion lattices as well. Monte Carlo simulations of the
microscopic spin model by Nishikawa et al. \citep{Nishikawa-PRB2019}
demonstrated a direct melting of the skyrmion lattice into a 2D liquid
with short-range correlations as in 3D and no evidence of the intermediate
hexatic phase. A similar conclusion was made in our recent Monte-Carlo
studies \citep{GC-PRB2023} of lattices of up to $10^{5}$ skyrmions
treated as particles with negative core energy and repulsive interaction
obtained from a microscopic spin model \citep{CGC-JPCM2020}. On the
contrary, simulations of skyrmion lattices in a GaV$_{4}$S$_{8}$
spinel by Baláž et al. \citep{Balaz-PRB2021} showed a two-step KTHNY
melting transition. Such a transition was also reported in experiments
on a Cu$_{2}$OSeO$_{3}$ nano-slab by Huang et al. \citep{Huang-Nat2020}
using cryo-Lorentz transmission electron microscopy.

In this paper, we compare the results of Monte Carlo studies of the
melting of a skyrmion lattice obtained within a point particles (PP)
model, as described above, and also within the underlying microscopic
lattice-spins (LS) model. Good agreement between the two models is
demonstrated and the single transition is confirmed. It exhibits a
kink in the magnetization that provides a convenient tool for experimental
detection of melting. We also explore a novel feature of the 2D melting
transition in skyrmion lattices that is absent in atomic systems.
Unlike a 2D atomic monolayer, the skyrmion lattice can exhibit melting
not just on temperature $T$ at a fixed magnetic field $H$ but also
on changing the field $H$ at a fixed temperature $T$. The $(T,H)$
phase diagram is computed, which can be tested in experiments. Moreover,
non-uniform ordering arises in the presence of the magnetic-field
gradient.

The paper is organized as follows. The theoretical background for
the studies of equilibrium skyrmion lattices and the degree of the
orientational order in the lattice is given in Sec.\ \ref{Sec_Theory}.
Numerical methods are described in Sec.\ \ref{Sec_numerical_method}.
The temperature dependence of the orientational order parameter is
computed in Sec.\ \ref{Sec_melting}. Section \ref{Sec_NS} shows
the results for the number of skyrmions in the system at different
temperatures, obtained within the LS model. The $(T,H)$ phase diagram
is presented and discussed in Sec.\ \ref{Sec_PD}. Non-uniform ordering
in the presence of the magnetic-field gradient is studied in Sec.
\ref{Sec_HGrad}. Finally, Sec.\ \ref{Sec_Conclusion} contains concluding
remarks and suggestions for the experiment.

\section{Theory}

\label{Sec_Theory}

\subsection{The model}

\label{Sec_model}

Our model describes ferromagnetically coupled three-component classical
spin vectors ${\bf s}_{i}$ of length $1$ on a square lattice with
the energy given by 
\begin{eqnarray}
\mathcal{H} & = & -\frac{1}{2}\sum_{ij}J_{ij}\mathbf{s}_{i}\cdot\mathbf{s}_{j}-H\sum_{i}s_{iz}\nonumber \\
 &  & -A\sum_{i}\left[(\mathbf{s}_{i}\times\mathbf{s}_{i+\delta_{x}})_{x}+(\mathbf{s}_{i}\times\mathbf{s}_{i+\delta_{y}})_{y}\right].\label{Hamiltonian}
\end{eqnarray}
The nearest neighbors exchange interaction of strength $J>0$ favors
ferromagnetic ordering and incorporates the actual length of the spin.
The stabilizing field $H$ in the second (Zeeman) term is applied
in the negative \textit{z}-direction, $H<0$. The third term in Eq.\ (\ref{Hamiltonian})
is the Dzyaloshinskii-Moriya interaction (DMI) of the Bloch type of
strength $A$, with $\delta_{x}$ and $\delta_{y}$ referring to the
next nearest lattice site in the positive $x$ or $y$ direction.
In this configuration, the dominant direction of spins is down and
that in the skyrmions is up. The DMI of the Néel type is described
by the term $(\mathbf{s}_{i}\times\mathbf{s}_{i+\delta_{x}})_{y}-(\mathbf{s}_{i}\times\mathbf{s}_{i+\delta_{y}})_{x}$
and leads to the same results.

The spin field in 2D is characterized by the topological charge \cite{belpol75jetpl}
\begin{equation}
Q=\int\frac{dxdy}{4\pi}\:{\bf s}\cdot\frac{\partial{\bf s}}{\partial x}\times\frac{\partial{\bf s}}{\partial y}\label{Q}
\end{equation}
that takes quantized values $Q=0,\pm1,\pm2,...$. For the pure-exchange
model, the analytical solution for topological configurations with
a given value of $Q$ was found by Belavin and Polyakov \cite{belpol75jetpl}
(see also Ref. \cite{CGC-JPCM2020}). Skyrmions and antiskyrmions
correspond to $Q=\pm1$. Due to the scale invariance of the exchange
interaction in 2D, their energy with respect to the uniform state,
$\Delta E_{\mathrm{BP}}=4\pi J$, does not depend on their size $\lambda$.
In a continuous spin-field model, the conservation of the topological
charge prevents skyrmions from decaying. Finite lattice spacing $a$
breaks this invariance by adding a term of the order $-\left(a/\lambda\right)^{2}$
to the energy, which leads to the skyrmion collapse \citep{CCG-PRB2012}.
This result was generalized for topological structures with any $Q$
in Ref. \cite{capgarchu19prr}.

In the presence of the DMI, only skyrmions, but not antiskyrmions,
exist. To find a single-skyrmion spin configuration numerically, one
can start with any state with $Q=1$ and perform energy minimization
as described in Ref. \citep{derchugar2018}. The energy of a single
skyrmion with respect to the uniform ferromagnetic state becomes negative
for $A$ large enough and $|H|$ small enough, making the uniform
state unstable to the creation of skyrmions. However, skyrmions compete
with the laminar domains. A stable skyrmion solution exists in the
field interval $H_{s}\leq|H|\leq H_{c}$. Below $H_{s}\simeq0.55A^{2}/J$
skyrmions become unstable against converting into a laminar domain
structure \citep{GC-PRB2023}. Above $H_{c}\simeq0.97A^{3/2}/J^{1/2}$
skyrmions collapse \citep{derchugar2018}. The ratio of these fields
is $H_{c}/H_{s}\simeq1.76\left(J/A\right)^{1/2}$. In practice, $A<J$,
thus providing a finite field range for the existence of skyrmions.

\subsection{Skyrmion-skyrmion interaction and the skyrmion-lattice parameter}

\label{Sec_interaction}

In sufficiently dense skyrmion lattices that we study here, the short-range
repulsion between skyrmions is dominating, whereas their dipole-dipole
repulsion can be neglected \cite{CGC-JPCM2020}. The former is given
by 
\begin{equation}
U(d)\simeq F\exp\left(-\frac{d}{\delta_{H}}\right),\qquad F\equiv60J\left(\frac{A^{2}}{JH}\right)^{2},\label{U_interaction}
\end{equation}
where $d$ is the distance between the skyrmions' centers and $\delta_{H}=a\sqrt{J/|H|}$
is the magnetic length. This formula offers an alternative (to the
LS model) way of building a skyrmion lattice as a system of particles
with a repulsive interaction (\ref{U_interaction}). In this paper,
we use both methods to describe the melting of skyrmion lattices and
compare the results with each other.

The number of skyrmions $N_{S}$ in the system at $T=0$ can be found
from the minimization of the total energy. For a triangular lattice
of skyrmions of the period $a_{S}$ and total area $S$ 
\begin{equation}
N_{S}=\frac{2}{\sqrt{3}}\frac{S}{a_{S}^{2}}.\label{NS_via_aS}
\end{equation}
Each skyrmion in the lattice interacts with its six nearest neighbors,
whereas interaction with further neighbors is negligibly small. The
energy per skyrmion is 
\begin{equation}
E_{0}=\varDelta E+3F\exp\left(-\frac{a_{S}}{\delta_{H}}\right),\label{E0_def}
\end{equation}
where $\Delta E<0$ is the skyrmion's core energy computed from the
LS model \cite{GC-PRB2023}. The equilibrium state at $T=0$ can be
obtained by the minimization of the total energy, $\mathcal{E}=N_{S}E_{0}$
with respect to $a_{S}$, as described in Ref.\ \citep{GC-PRB2023}. 

For $A/J=0.2$ (that is used throughout the paper) and $H/J=-0.025$
one has $\Delta E=-4.23J$, $\delta_{H}=6.32a$, and $F=154J$ that
results in $a_{S}=38.5a$. The interaction energy between nearest
neighbors $U_{0}=F\exp\left(-a_{S}/\delta_{H}\right)=0.349J$ defines
the scale for the melting temperature that appears to be $T_{m}\simeq0.12J$. 

One can extend the theory, at least qualitatively, by minimizing the
free energy $\mathcal{F}=\mathcal{E}-T\mathcal{S}$, where $\mathcal{S}$
is the entropy,
\begin{equation}
\mathcal{S}=N_{S}\ln\mathcal{N}_{\mathrm{positions}},
\end{equation}
and $\mathcal{N}_{\mathrm{positions}}$ is the number of positions
available for skyrmions taking into account their repulsion. At the
temperature $T$ the minimal distance $r_{T}$ between skyrmions can
be estimated as $Fe^{-r_{T}/\delta_{H}}\simeq T$ that results in
\begin{equation}
r_{T}\simeq\delta_{H}\ln\frac{F}{T},\qquad\mathcal{N}_{\mathrm{positions}}\simeq\frac{S}{r_{T}^{2}}=\frac{S}{\delta_{H}^{2}}\frac{1}{\ln\left(F/T\right)}.
\end{equation}
Substituting this into the free energy $\mathcal{F}$ and using Eq.
(\ref{NS_via_aS}), one obtains
\begin{equation}
\mathcal{F}=\frac{2}{\sqrt{3}}\frac{S}{a_{S}^{2}}\left[\varDelta E-T\ln\left(\frac{S}{\delta_{H}^{2}}\frac{1}{\ln\left(F/T\right)}\right)+3F\exp\left(-\frac{a_{S}}{\delta_{H}}\right)\right].\label{Free_energy_final}
\end{equation}
This can be minimized numerically with respect to $a_{S}$, then the
number of skyrmions can be found from Eq. (\ref{NS_via_aS}). The
resulting dependence $N_{S}(T)$ shown in black in Fig. \ref{Fig_NS}
below is very close to a straight line. 

\subsection{Skyrmion lattice}

\label{lattice}

\begin{figure}
\begin{centering}
\includegraphics[width=8cm]{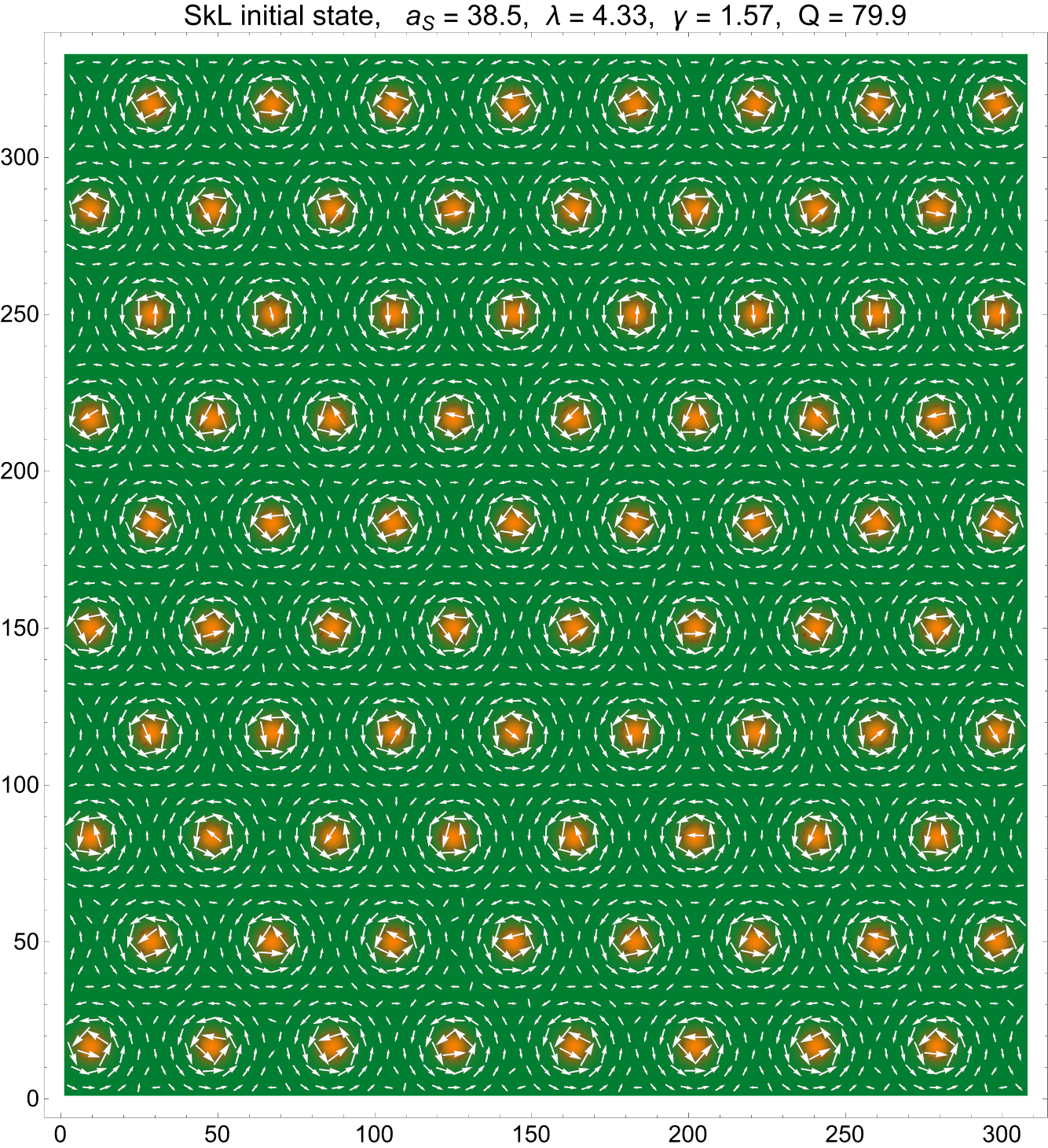}
\par\end{centering}
\caption{Initial state for the triangular lattice of skyrmions. Spin components
are color-coded, $s_{z}=-1$ green, $s_{z}=1$ orange. White arrows
show the in-plane spin components $s_{x}$ and $s_{y}$.}
\label{Fig-LIC} 
\end{figure}

\begin{figure}
\begin{centering}
\includegraphics[width=8cm]{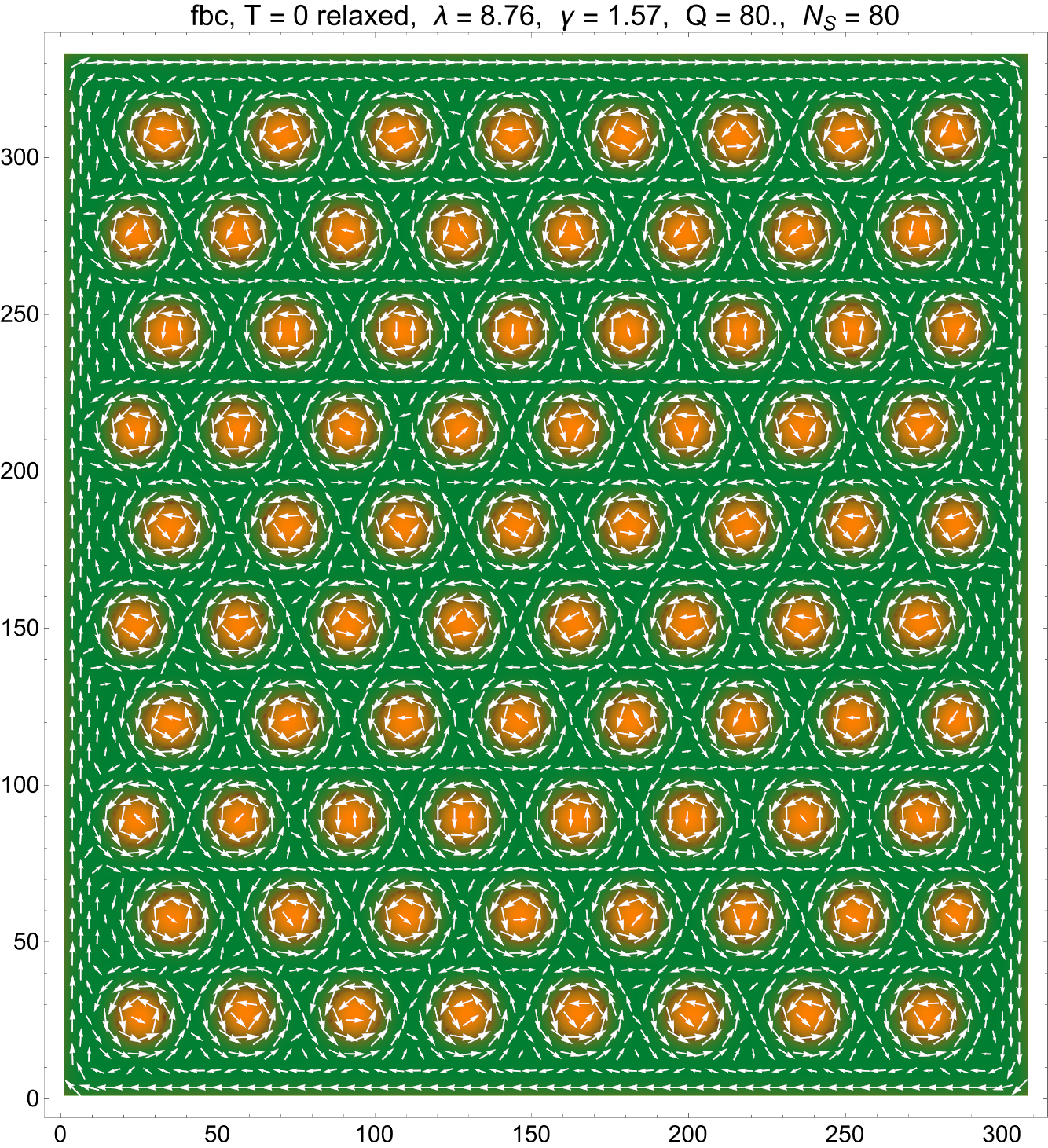}
\par\end{centering}
\caption{Skyrmion lattice in the lattice-spin model with free boundary conditions
obtained through relaxation of the initial state of bubbles with $Q=1$,
Fig. \ref{Fig-LIC}.}

\label{Fig_SkL_fbc}
\end{figure}

The positions of skyrmions in a triangular lattice are given by 
\begin{equation}
\frac{{\bf R}}{a_{S}}={\bf e}_{x}n_{x}+\mathbf{e}_{60}n_{60}={\bf e}_{x}\left(n_{x}+\frac{1}{2}n_{60}\right)+{\bf e}_{y}\frac{\sqrt{3}}{2}n_{60},\label{Triangular_lattice_positions}
\end{equation}
where ${\bf e}_{x}$ and ${\bf e}_{y}$ are unit vectors along $x,y$
coordinate axes, $\mathbf{e}_{60}=(1/2){\bf e}_{x}+(\sqrt{3}/2){\bf e}_{y}$
is the lattice vector directed at $60{^\circ}$ to the $x$-axis and
$n_{x}$, $n_{60}$, are integers. We use the value of $a_{S}$ obtained
by the energy minimization. The regions of $n_{x},n_{60}$ and the
dimensions of the system are chosen to avoid distortions of the lattice
near the boundaries in the case of periodic boundary conditions (pbc).
An example of the initial state used for obtaining the skyrmion lattice
within the model of lattice spins is shown in Fig. \ref{Fig-LIC}.
It consists of Bloch-type bubbles placed in positions given by Eq.
\ref{Triangular_lattice_positions}. Energy minimization at $T=0$
results in the skyrmion lattice that for the model with pbc looks
similar, except that the size of skyrmions increases {[}same as for
the model with free boundary conditions (pbc) below{]}. The results
for the energy minimization for the fbc model are shown in Fig. \ref{Fig_SkL_fbc}.
Because of the repulsion from the boundary, the lattice is somewhat
compressed and distorted near the vertical walls. This effect becomes
non-essential in systems of a larger size. 

For the model of point particles with repulsion, the particles are
initially placed at the positions according to Eq. (\ref{Triangular_lattice_positions})
and then Monte Carlo simulation is performed. For $T=0$, this leads
to the energy minimization in general. For the pbc model in this case,
the particles remain at their initial positions. The result for the
rigid-wall condition at $T=0$ is similar to that for the LS model
with fbc. 

\subsection{Orientational order}

\label{Sec_Orientational_order}

The quantity describing the orientation of the hexagon of the nearest
neighbors of any skyrmion $i$ in a 2D lattice is local hexagonality
\begin{equation}
\Psi_{i}=\frac{1}{6}\sum_{j}\exp(6i\theta_{ij}).\label{Hexagonality_def}
\end{equation}
Here the summation is carried out over 6 nearest neighbors $j$, $\theta_{ij}$
is the angle between the $ij$ bond and any fixed direction in the
lattice. If $\theta$ is counted from the direction of the $x$-axis
(that is our choice), and two of the sides of the perfect hexagon
coincide with the $x$-axis (as in the figures above), all terms in
the sum are equal to $1$, and thus $\Psi_{i}=1$. We call this \textit{horizontal}
orientation of hexagons. Rotation of the hexagons by 30$^{\circ}$
from the horizontal orientation results in the \textit{vertical} orientation
for which $\Psi_{i}=-1$. For any other orientation of a perfect hexagon,
$\Psi_{i}$ is a complex number of modulus 1. The angle $\phi_{i}$
by which the hexagon $i$ is rotated from its initial horizontal orientation
is related to the phase angle $\Theta_{i}$ in $\Psi_{i}=\left|\Psi_{i}\right|e^{i\Theta_{i}}$
as $\phi_{i}=\Theta_{i}/6$.

At finite temperatures, orientations of the bonds fluctuate and the
condition $|\Psi_{i}|=1$ no longer holds. The quality of hexagons
can be described by global hexagonality value \citep{GC-PRB2023}
defined as
\begin{equation}
V_{6}=\sqrt{\frac{1}{N_{S}}\sum_{i}|\Psi_{i}|^{2}},\label{V6_def}
\end{equation}
where $N_{S}$ is the number of skyrmions in the system. At temperatures
well above the melting transition, when even the short-range order
is completely destroyed, the orientations of the bonds and the angles
$\theta_{ij}$ become random. In this limit $V_{6}=\sqrt{1/6}$ as
each particle has six nearest neighbors on average. The common orientation
of hexagons in the lattice is described by the complex order parameter
which is defined as local hexagonality averaged over the system:
\begin{equation}
O_{6}=\frac{1}{N_{S}}\sum_{i}\Psi_{i}.\label{O6_def}
\end{equation}

\section{Numerical method}

\label{Sec_numerical_method}

The lattice-spin model was numerically solved with the help of the
energy minimization at $T=0$ and by the Metropolis Monte Carlo routine
at $T>0$, as described in Ref. \cite{GCZZ-MMM2021}. At $T=0$ we
applied sequential rotation of spins $\mathbf{s}_{i}$ toward their
effective field $\mathbf{H}_{\mathrm{eff},i}=-\partial\mathcal{H}/\partial\mathbf{s}_{i}$
with the probability $\alpha$ and the energy-conserving overrelaxation
(rotating the spins by 180$^{\circ}$ around $\mathbf{H}_{\mathrm{eff},i}$)
with the probability $1-\alpha$. At $T>0$, we combined the Monte
Carlo update and overrelaxation with the same probabilities. In most
computations we used $\alpha=0.01$ which allows efficient exploring
of the phase space of the system by overrelaxation. Monte Carlo updating
of all spins in the system constitutes one Monte Carlo step (MCS).
We used both lattice initial condition (LIC) and random initial condition
(RIC) in which the directions of lattice spins are random. The latter
is used to study the freezing of the skyrmion lattice that for large
systems leads to a polycrystalline state as different regions freeze
with different directions of hexagons and the motion of domain boundaries
is very slow below the freezing transition.

We studied the systems of a nearly square shape with dimensions adjusted
to the skyrmion lattice with the lattice parameter $a_{S}$ computed
by the energy minimization as explained above. At the boundaries,
we used periodic boundary conditions (pbc) and free boundary conditions
(fbc).

\begin{figure}
\begin{centering}
\includegraphics[width=8cm]{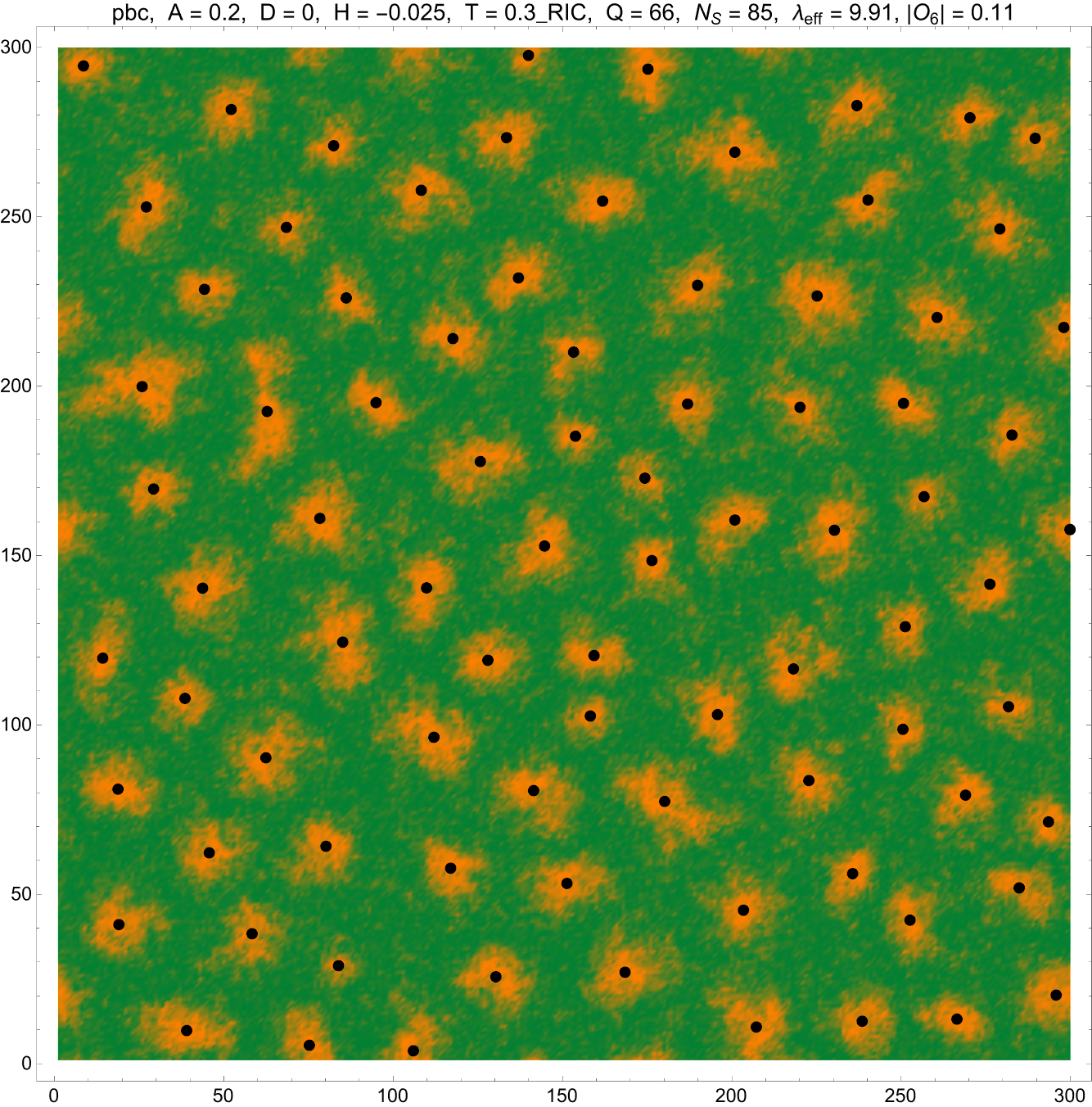}
\par\end{centering}
\caption{A snapshot of the skyrmion liquid at $T/J=0.3$ (above the skyrmion-lattice
melting point). Thermal agitation washes out skyrmions' shape. Black
points are skyrmions' positions found by the skyrmion locator, Eq.
(\ref{Skyrmion_locator}).}

\label{Fig_Skyrmion_location}
\end{figure}

To identify skyrmions, we used an adaptation of the well-known algorithm
for labeling islands using, as parameters, the \textit{sea level}
(the value of $s_{z}$ identifying the boundary of the skyrmion's
core) equal to zero, \textit{connection range} (the maximal distance
between the spins along the $x$ and $y$ axes for which they can
be considered as neighbors) equal to $a$ (so that each spin has 8
neighbors), and the \textit{minimal island size} (the number of spins
with $s_{z}>0$ in the skyrmion) equal to five. This algorithm checks
all spins in the lattice for $s_{z}>0$. If such spin is found, all
its neighbors within the connection range are checked and added to
the island if they have $s_{z}>0$. This ends when there are no more
connected neighbors with $s_{z}>0$. If the number of spins within
the island is greater or equal to the minimal island size, the skyrmion
is identified. The limitation on the skyrmion size is important at
elevated temperatures when spins are substantially disordered and
there can be even isolated spins pointing up that should not be counted
as skyrmions. After all skyrmions are identified, their centers $\mathbf{R}_{i}$
can be found with the help of the skyrmion-locator formula
\begin{equation}
\mathbf{R}_{i}=\sum_{j\in i}\mathbf{r}_{j}s_{z,j}^{2}/\sum_{j\in i}s_{z,j}^{2},\label{Skyrmion_locator}
\end{equation}
where $j\in i$ are all lattice sites that belong to the skyrmion
$i$. Here the weight factor $s_{z,j}^{2}$ favors the sites close
to the skyrmion's top. An example of the skyrmion location is shown
in Fig. \ref{Fig_Skyrmion_location}. One can see that at elevated
temperatures well above melting, skyrmions are strongly washed out
by thermal agitation. 

After skyrmions are identified, the system is subdivided into square
bins containing 4-6 skyrmions. In the computation of the orientational
order, Eqs. (\ref{Hexagonality_def})-(\ref{O6_def}), the six nearest
neighbors are searched for only within the same bin and the eight
surrounding bins. This greatly speeds up the computation.

Computations were performed with Wolfram Mathematica using compilation
of the core routines such as the energy minimization and Monte Carlo.
We used versions with and without internal parallelization. In the
latter, the system is split into the intermittent ABABAB... stripes
the number of each of them being equal to the number of used processor
cores $N_{\mathrm{cores}}$. First, all A-stripes are processed in
parallel, then all B-stripes are processed in parallel, and so on.
The faster parallelized version was used to study the evolution and
the final state of the system at one temperature. To obtain temperature
dependences, we used the non-parallelized version computing different
temperature points in a parallelized cycle. This works faster than
the parallelized version because there is no overhead due to internal
parallelization. 

The parallelized-cycle computations were done in the so-called \textit{relay
}mode in which the next $T$-point within the same thread used the
previously obtained state as the initial condition. With the interval
between the neighboring $T$-points being $\delta T$, that between
those within the same thread was $\Delta T=N_{\mathrm{cores}}\delta T$
that is sufficiently small if $\delta T$ is small and $N_{\mathrm{cores}}$
is not too large.

The largest spin system considered here is about $1155\times1200$
lattice sites which is more than one million spins. For our set of
parameters, the perfect skyrmion lattice in this system contains only
1080 skyrmions which is sufficient to observe the melting of the skyrmion
lattice and other effects. However, computations with such a large
number of spins take a lot of computer time and it is difficult to
scale up the system.

Considering skyrmions as point particles allows performing computations
on much larger systems. In Ref. \cite{GC-PRB2023} Monte Carlo simulations
were done on the systems of up to $10^{5}$ particles that showed
a first-order phase transition with a narrow $T$-hysteresis loop.
Here, to compare the LS and PP models, it is sufficient to simulate
the systems of $10^{4}$ particles. The Metropolis Monte Carlo routine
makes successive trial displacements of particles and the trials are
accepted if $\mathrm{rand}<\exp\left(-\Delta E/T\right)$, where rand
is a random number in the interval (0,1) and $\Delta E$ is the energy
change associated with the trial displacement. The average value of
the trial displacement was programmed auto-adjusting to maximize the
effective displacement (the average trial displacement times the acceptance
rate) and limited from above by $0.3a_{S}$. As in Ref. \cite{GC-PRB2023},
in computing the repulsion energy, we used the distance cutoff $r_{\mathrm{cutoff}}=0.95\sqrt{3}a_{S}$
that is just shy of the next-nearest distance in the triangular lattice,
$\sqrt{3}a_{S}$. The system was subdivided into square bins of a
size slightly larger than $r_{\mathrm{cutoff}}$ so that any particle
in the bin could interact with those in the same and eight surrounding
bins. With the number of particles 4-6 in the bin, this severely limits
the number of interacting partners to consider and greatly speeds
up the simulation.

In the Monte Carlo simulation of both LS and PP models, we used the
stopping criterion requiring the derivative of a control quantity
(CQ) with respect to MCS to be small enough. The CQ was the energy
$U$ per particle in the case of the PP model. For the LS model, we
used $V_{6}$ given by Eq. (\ref{V6_def}) as CQ, to put the focus
on skyrmions rather than on spins. We averaged the CQ over the last
30\% of all MCS' and required $\Delta\left\langle \mathrm{CQ}\right\rangle /(\Delta\mathrm{MCS)}<CQ_{\mathrm{scale}}/\mathrm{MCSens}$,
where $CQ_{\mathrm{scale}}$ is the scale of the CQ: the interaction
energy $U_{0}=F\exp\left(-a_{S}/\delta_{H}\right)$ for $U$ and 1
for $V_{6}$, while MCSens is the sensitivity of the Monte Carlo routine,
$\mathrm{MCSens}=10^{-7}$ for most simulations. Additionally, we
required that the stopping criterion is fulfilled in total for $10^{4}$
MCS to exclude a premature stopping in the case the CQ increases and
decreases because of fluctuations. Other computed quantities such
as $V_{6}$ and $O_{6}$ were averaged over the last 30\% of all MCS',
similar to the control quantity. The simulations were performed in
blocks of 20 or 50 MCS, to reduce the overhead of computing the system
energy or hexagon quality. The number of MCS before stopping was above
$10^{5}$ in the vicinity of the melting-transition temperature $T_{m}$,
smaller above $T_{m}$ and even smaller well below $T_{m}$.

Computations were performed on a 20-core Dell Precision workstation
with 16 cores licensed for Mathematica, as well as on smaller machines.

\section{Numerical results}

\label{Sec_Numerical_results}

\subsection{Melting of the skyrmion lattice}

\label{Sec_melting}

\begin{figure}
\begin{centering}
\includegraphics[width=8cm]{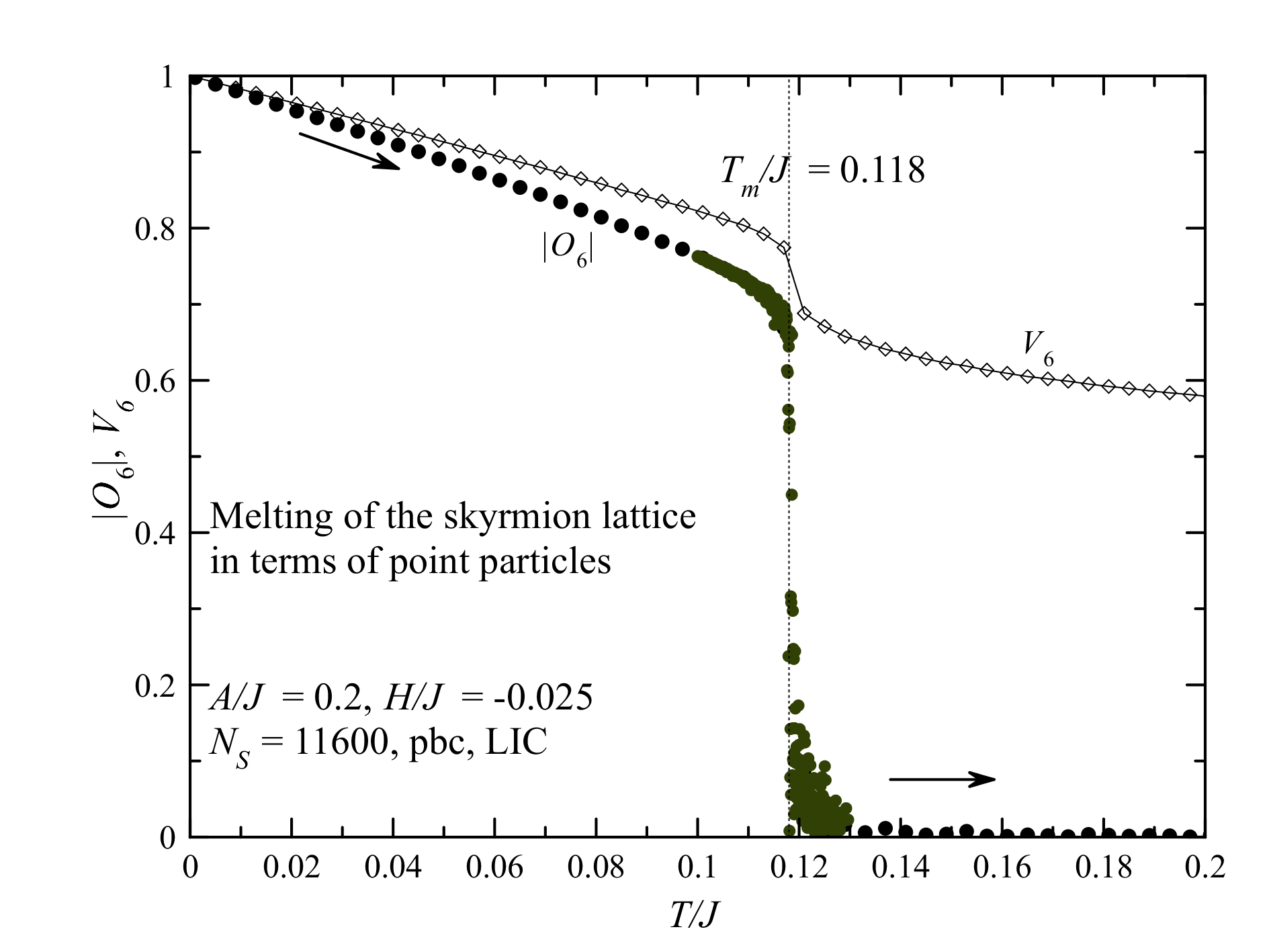}
\par\end{centering}
\caption{Melting of the skyrmion lattice with periodic boundary conditions
in terms of point particles, starting at $T=0$ with LIC.}

\label{Fig_O5V6_PP}
\end{figure}

\begin{figure}
\begin{centering}
\includegraphics[width=8cm]{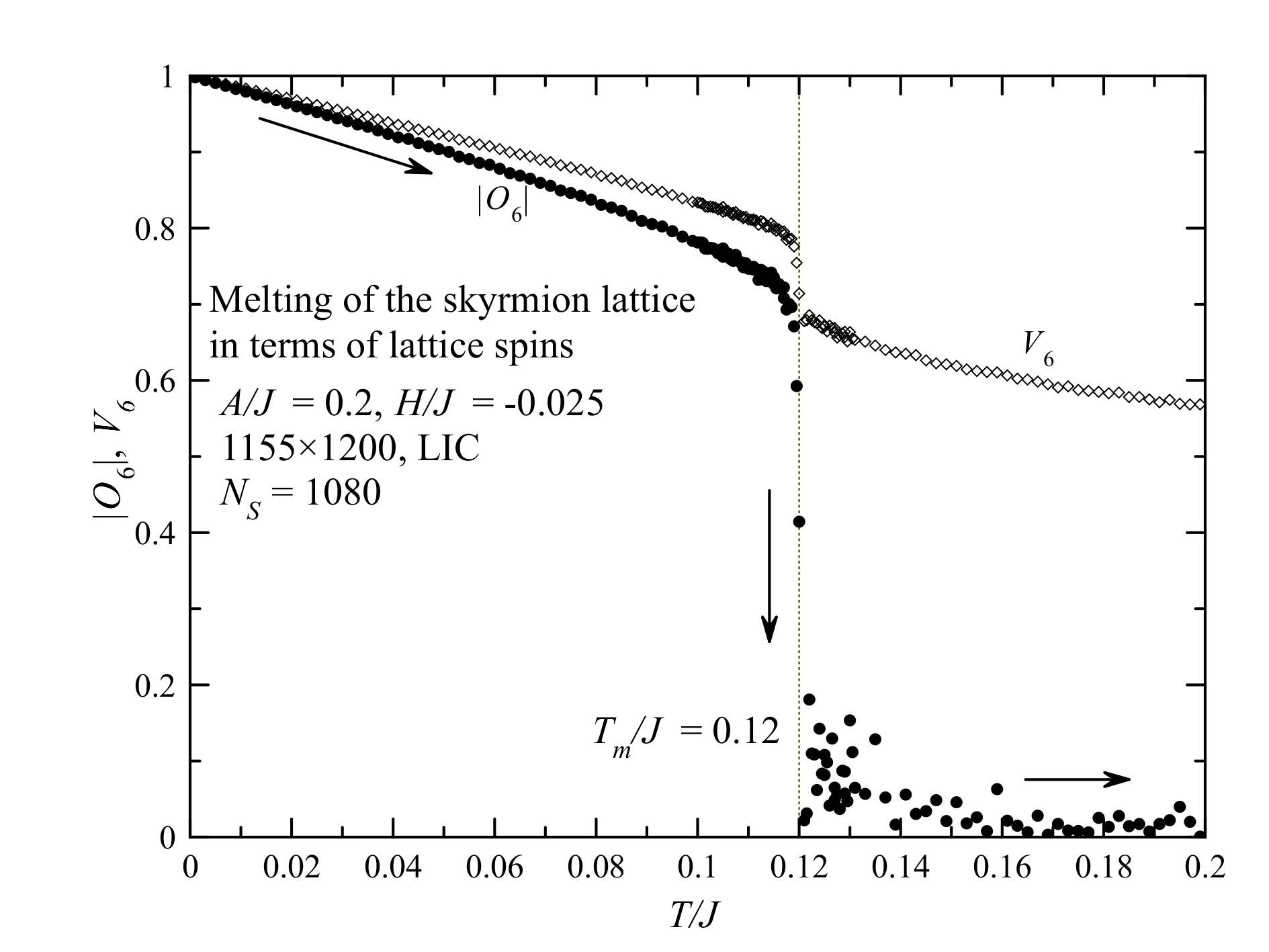}
\par\end{centering}
\caption{Melting of the skyrmion lattice with periodic boundary conditions
in terms of lattice spins, starting at $T=0$ with LIC.}

\label{Fig_O6V6_LS}
\end{figure}

The temperature dependences of the orientational order parameter $O_{6}$
and the hexagonality value $V_{6}$ for the system of $10^{4}$ skyrmions
within the point-particle model are shown in Fig. \ref{Fig_O5V6_PP}.
The melting occurs as a single transition at $T_{m}=0.118J$ as in
Ref. \cite{GC-PRB2023} where a larger system of $10^{5}$ particles
was simulated. The melting transition is likely first order, although
further efforts are needed to fully clarify the critical behavior.
The results for the lattice spins model with pbc shown in Fig. \ref{Fig_O6V6_LS}
are similar, and there is an excellent agreement between the values
of $T_{m}$ obtained within the two models, despite the number of
skyrmions in the LS model is only $10^{3}$. 

\begin{figure}
\begin{centering}
\includegraphics[width=8cm]{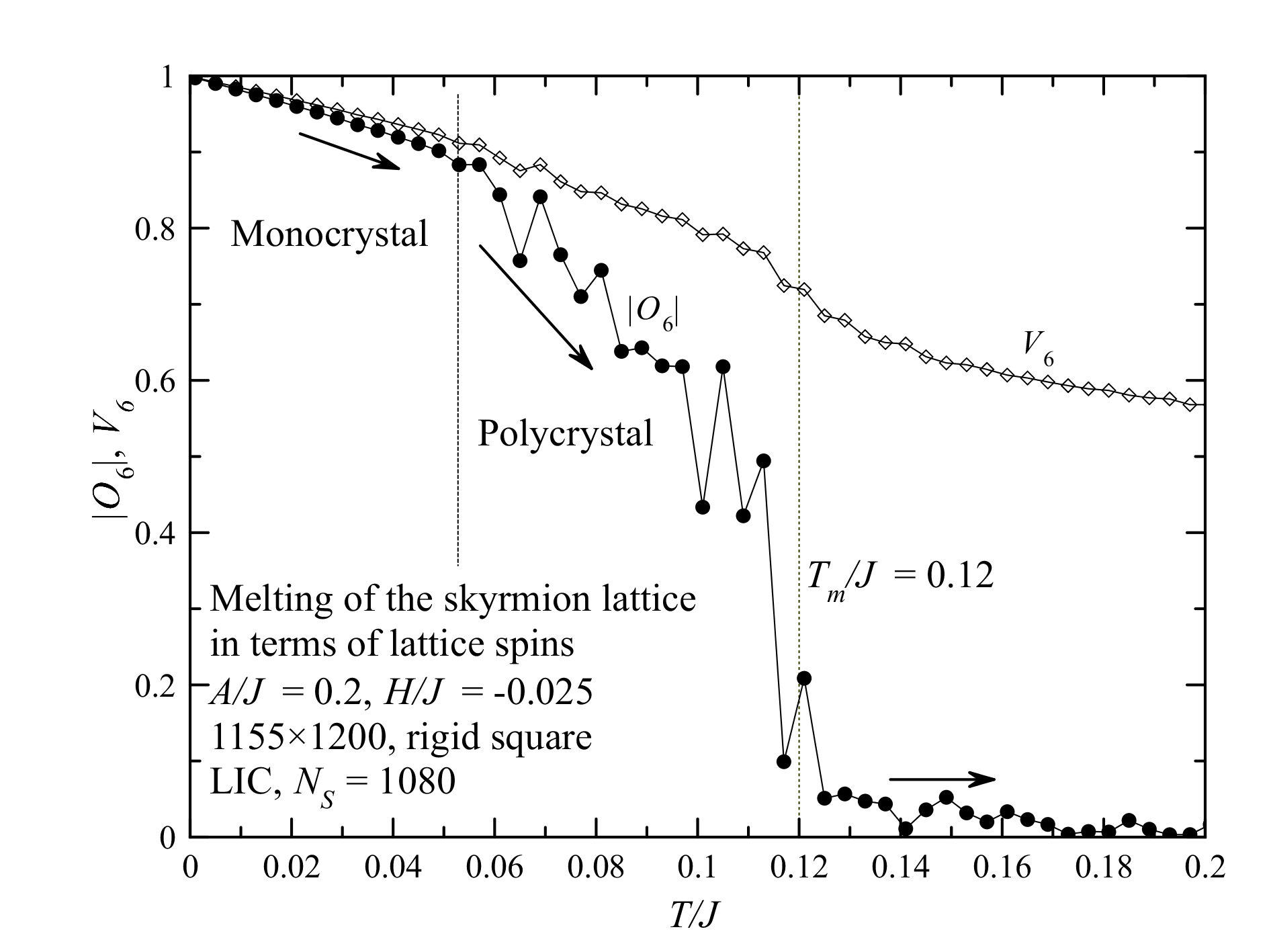}
\par\end{centering}
\caption{Melting of the skyrmion lattice with free boundary conditions in terms
of lattice spins, starting at $T=0$ with LIC. Flat boundaries favor
the orientation of hexagons parallel to the boundary. This results
into breaking an initially monocrystalline skyrmion lattice into a
polycrystalline state with the orientation of hexagons defined by
the closest boundary. This leads to an irregular dependence $|O_{6}(T)|$.}

\label{Fig_O6V6_LS_fbc}
\end{figure}

\begin{figure}
\begin{centering}
\includegraphics[width=8cm]{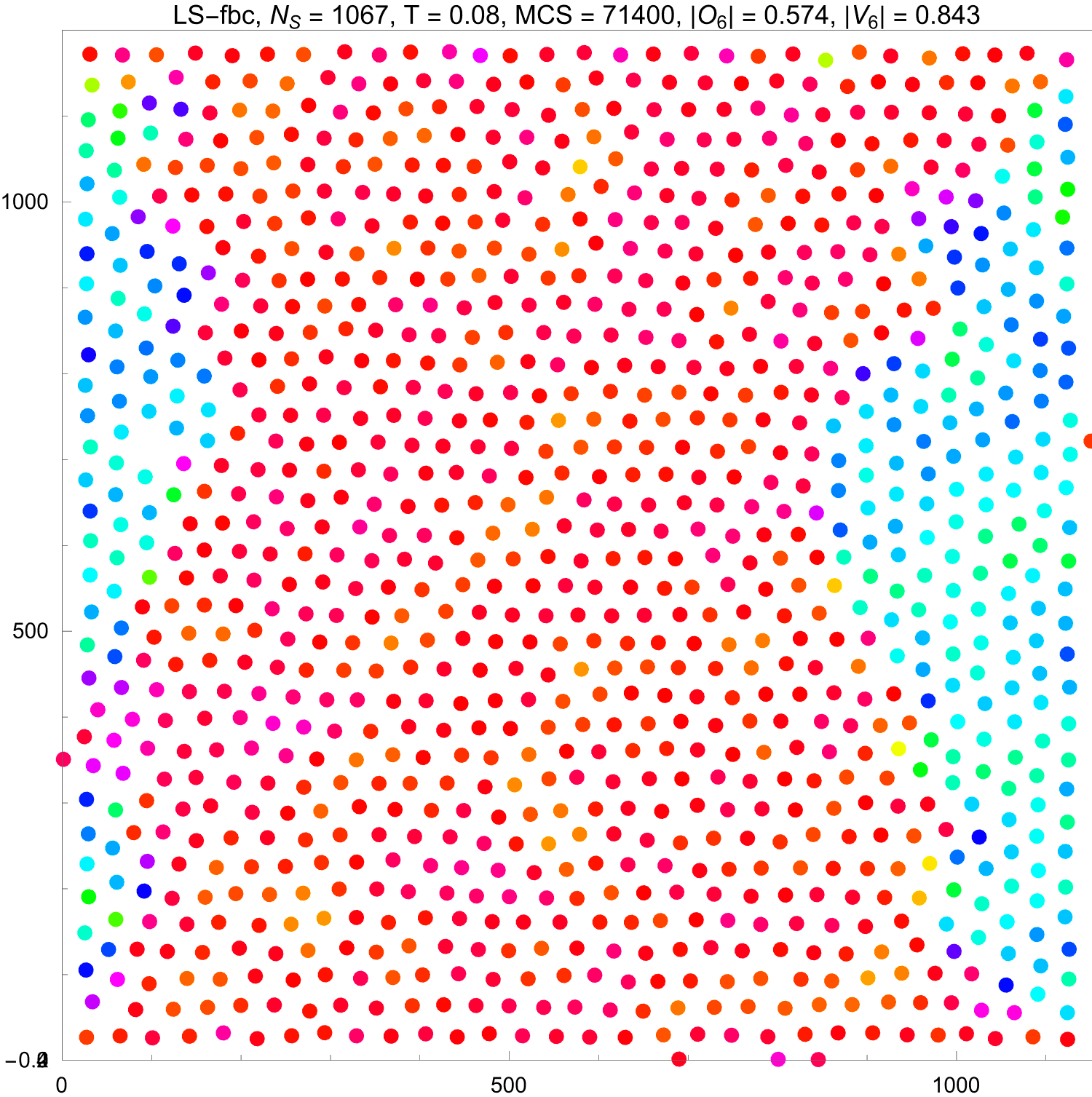}
\par\end{centering}
\caption{The polycrystal state of the skyrmion lattice in the lattice-spins
system with free boundary conditions at $T/J=0.08$. Boundaries favor
the orientation of hexagons (color coded) parallel to the boundary.
The initial state is a lattice with the horizontal hexagon orientation
(red).}

\label{Fig_Polycrystal_lattice}
\end{figure}

Free boundary conditions lead to the breakdown of the skyrmion lattice
well below the melting transition, as can be seen in Fig. \ref{Fig_O6V6_LS_fbc}
showing the results for the LS model. As was observed in Ref. \cite{GC-PRB2023},
boundaries polarize the skyrmion solid favoring the orientation of
hexagons parallel to the boundary. This effect is the strongest for
the system of rhomboid shape in which all boundaries work in the same
direction which results in a significant finite-size effect inducing
a non-zero $O_{6}$ above the melting point. The system of a square
shape is frustrated since the horizontal and vertical boundaries work
in different directions. As a result, the initially prepared perfect
skyrmion lattice with increasing temperature breaks up into a polycrystal
in which the boundaries dictate the orientation of hexagons in their
vicinity, see Fig. \ref{Fig_Polycrystal_lattice}. Similar results
are obtained for the PP system of a square shape with the rigid-wall
boundary conditions.


\subsection{The magnetization anomaly}

\label{Sec_Magnetization-anomaly}

\begin{figure}
\begin{centering}
\includegraphics[width=8cm]{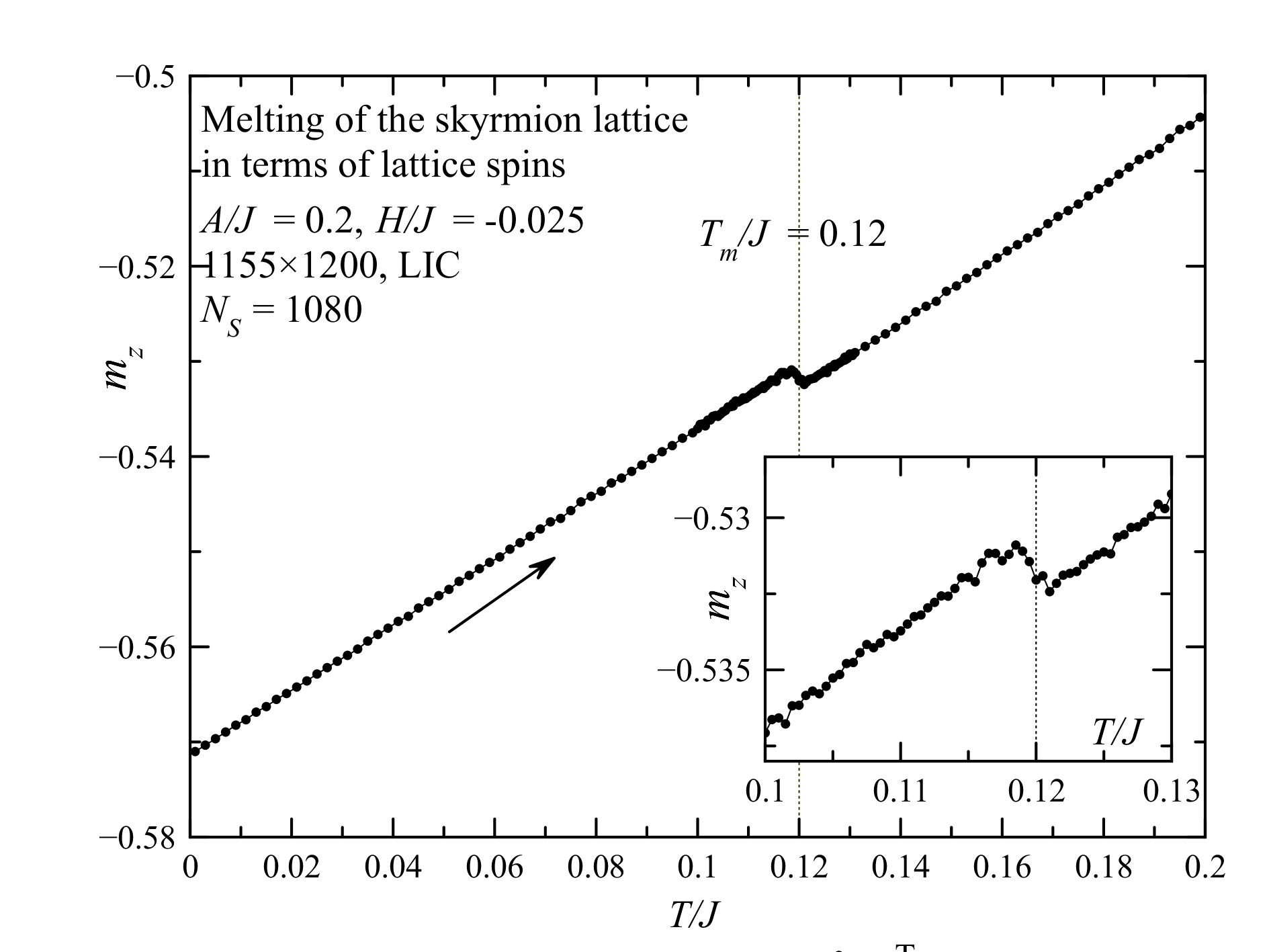}
\par\end{centering}
\caption{Temperature dependence of the magnetization shows a kink at the melting
point.}

\label{Fig_mz_vs_T}
\end{figure}
As we have seen, most of the results about the skyrmion lattice can
be obtained from the point-particle model which is computationally
less demanding than the model of lattice spins and allows simulation
of larger systems. However, the LS model shows an interesting kink
anomaly of the magnetization at the melting point that can be used
to locate the melting transition as the magnetization can be measured
with great precision. Fig. \ref{Fig_mz_vs_T} shows an expected linear
decrease of the magnitude of the magnetization $\mathbf{m}=(1/N)\sum_{i}\left\langle \mathbf{s}_{i}\right\rangle $
with increasing temperature and, on the top of it, a sudden increase
of it in the narrow region of melting.

\subsection{The number of skyrmions}

\label{Sec_NS}

\begin{figure}
\begin{centering}
\includegraphics[width=8cm]{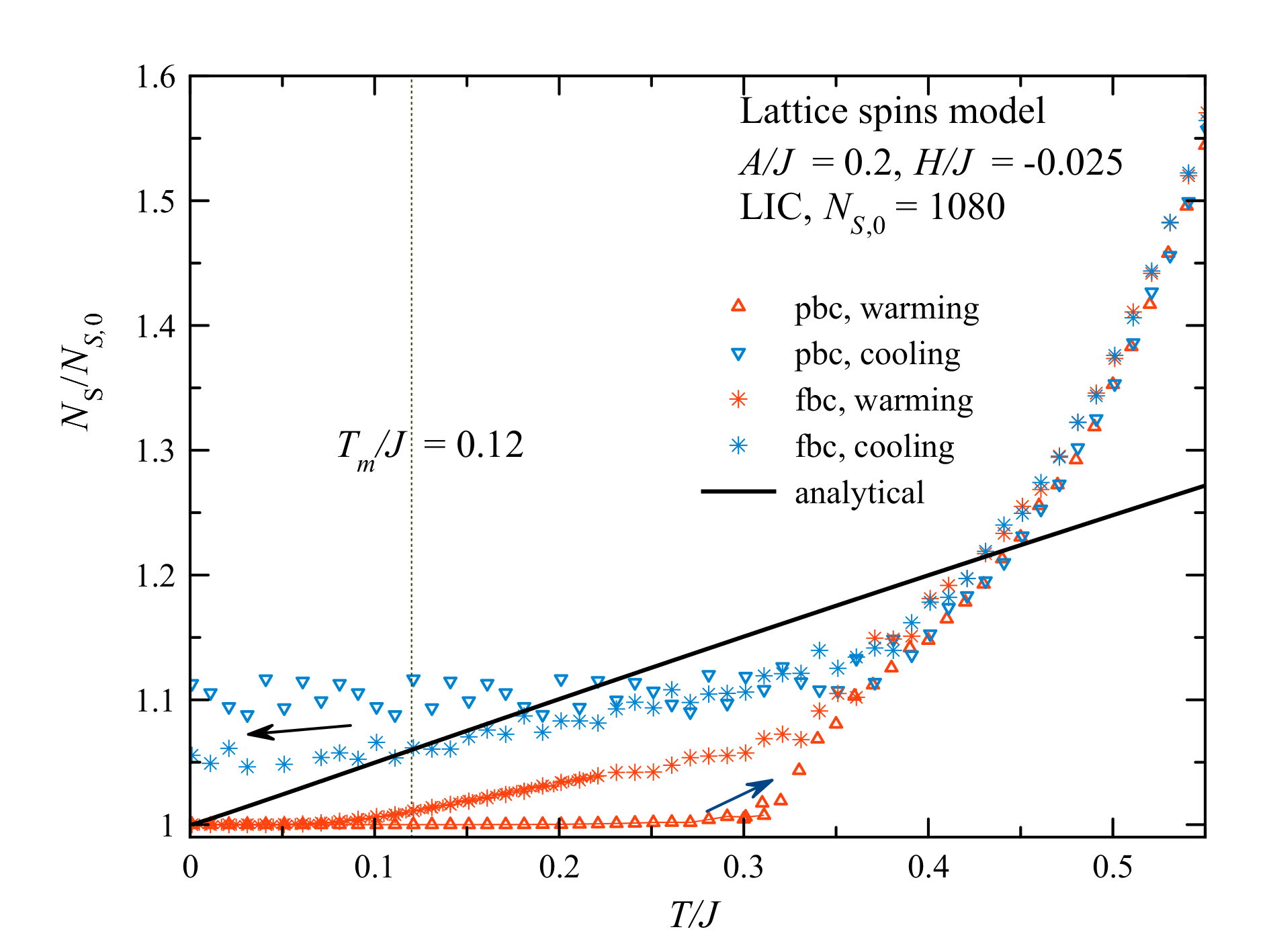}
\par\end{centering}
\caption{Temperature dependence of the number of skyrmions in the model of
lattice spins. In the pbc model, $N_{S}$ barely changes below $T/J=0.3$.
In the fbc model, there is some increase due to the creation of skyrmions
in the cracks between crystallites on the polycrystalline state. Both
dependences have a hysteresis. The black line in the result of the
numerical minimization of the free energy, Eq. (\ref{Free_energy_final}).}

\label{Fig_NS}
\end{figure}
The equilibrium number of skyrmions $N_{S}$ discussed in Sec. \ref{Sec_interaction}
is, in fact, a complicated issue because at low temperatures $N_{S}$
is very difficult to change. Creation of a skyrmion requires overcoming
an energy barrier only slightly lower than $4\pi J$ (see Ref. \cite{derchugar2018}
for details). Annihilation of a skyrmion with a negative energy $\Delta E$
requires overcoming of an even larger barrier. In the region of the
melting, $T\sim0.1J$, the probability of these processes is exponentially
small which makes them hardly accessible to Monte Carlo simulations.
In the experiment, however, the situation can be different because
there is a huge factor between real time and computer time so an exceedingly
long computation can correspond to a short real time. In any case,
changing of the number of skyrmions in the system is a slow process
because within the continuous approximation, skyrmions are topologically
protected and the topological charge $Q$ of the system can change
only via the lattice-mediated processes \cite{CCG-PRB2012}.

Treating skyrmions as point particles in Monte Carlo simulations allows
for bypassing the barrier of the skyrmion creation/annihilation. However,
for our set of parameters, removing a skyrmion from a lattice costs
the energy $2.14J$ while adding a skyrmion into a lattice lacuna
costs the energy $9.89J$ \cite{GC-PRB2023}. Although these energies
are smaller than $4\pi J$, they are large enough to prevent equilibration
of the number of skyrmions below melting with the help of a Monte
Carlo simulation, if it were extended to include C/A processes. 

The results obtained for the lattice-spins model with pbc (in which
the number of skyrmions is not fixed and can change naturally) are
shown in Fig. \ref{Fig_NS}. For the model with pbc containing $N_{S,0}=1080$
skyrmions in LIC at $T=0$, on increasing the temperature $N_{S}$
starts to grow only at $T/J\simeq0.3$ and then goes up steeply. On
cooling, $N_{S}$ does not fully return to its initial value, so that
$N_{S}/N_{S.0}\simeq1.1$ at low temperatures. The number of skyrmions
becomes non-equilibrium below $T/J\simeq0.4$ in this simulation.

For the LS model with fbc the results are similar except for the growth
of $N_{S}$ that starts already at $T/J\simeq0.1$. This can be attributed
to the boundaries between domains with different orientations of hexagons
in the polycrystal lattice (see Fig. \ref{Fig_Polycrystal_lattice})
where the lattice is broken and it should be easier to create or annihilate
skyrmions. The massive growth of $N_{S}$ starts, however, only at
$T/J\simeq0.3$, as in the model with pbc. The fbc model also shows
a hysteresis, although weaker than the pbc model.

In Fig. \ref{Fig_NS}, the analytical result for the number of skyrmions
following from the minimization of the free energy, Eq. (\ref{Free_energy_final})
is shown by the black line that is almost straight and different from
the numerical results for the LS model with pbc and fbc. The lack
of agreement can be explained by the two factors. First, Monte Carlo
results at low temperatures are non-equilibrium with respect to $N_{S}$.
Second, at elevated temperatures, the model of point particles cannot
be quantitatively correct because skyrmions are strongly distorted
by thermal fluctuations (see Fig. \ref{Fig_Skyrmion_location}).

\subsection{Phase diagram}

\label{Sec_PD}

\begin{figure}
\begin{centering}
\includegraphics[width=8cm]{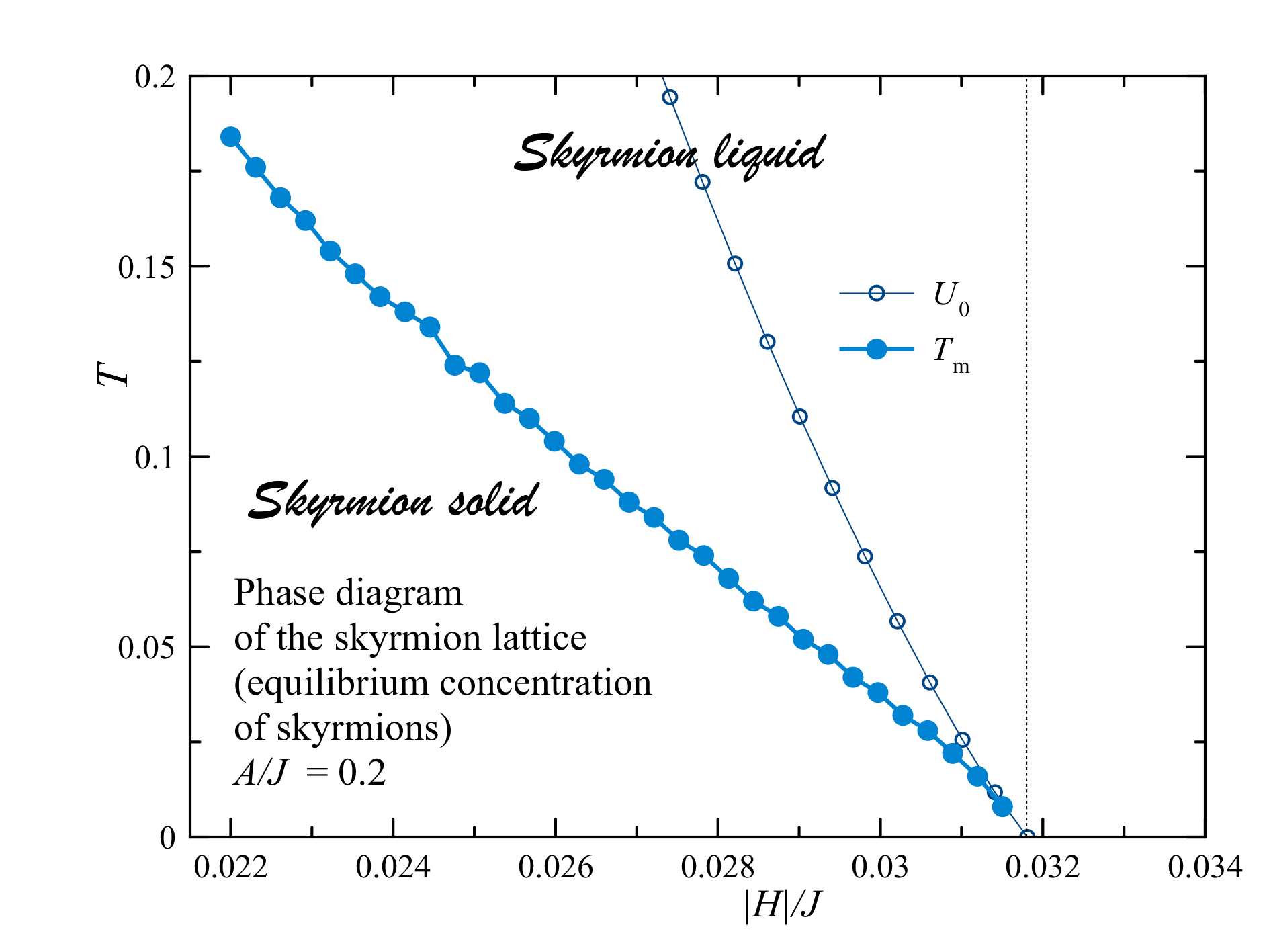}
\par\end{centering}
\caption{The phase diagram of the skyrmion liquid-solid system assuming the
concentration of skyrmions equal to their equilibrium concentration
at $T=0$ for any value of $H$. The melting temperature correlates
with the skyrmion-skyrmion repulsion energy $U_{0}$ that is also
shown.}

\label{Fig_PD_equilibrium}
\end{figure}

\begin{figure}
\begin{centering}
\includegraphics[width=8cm]{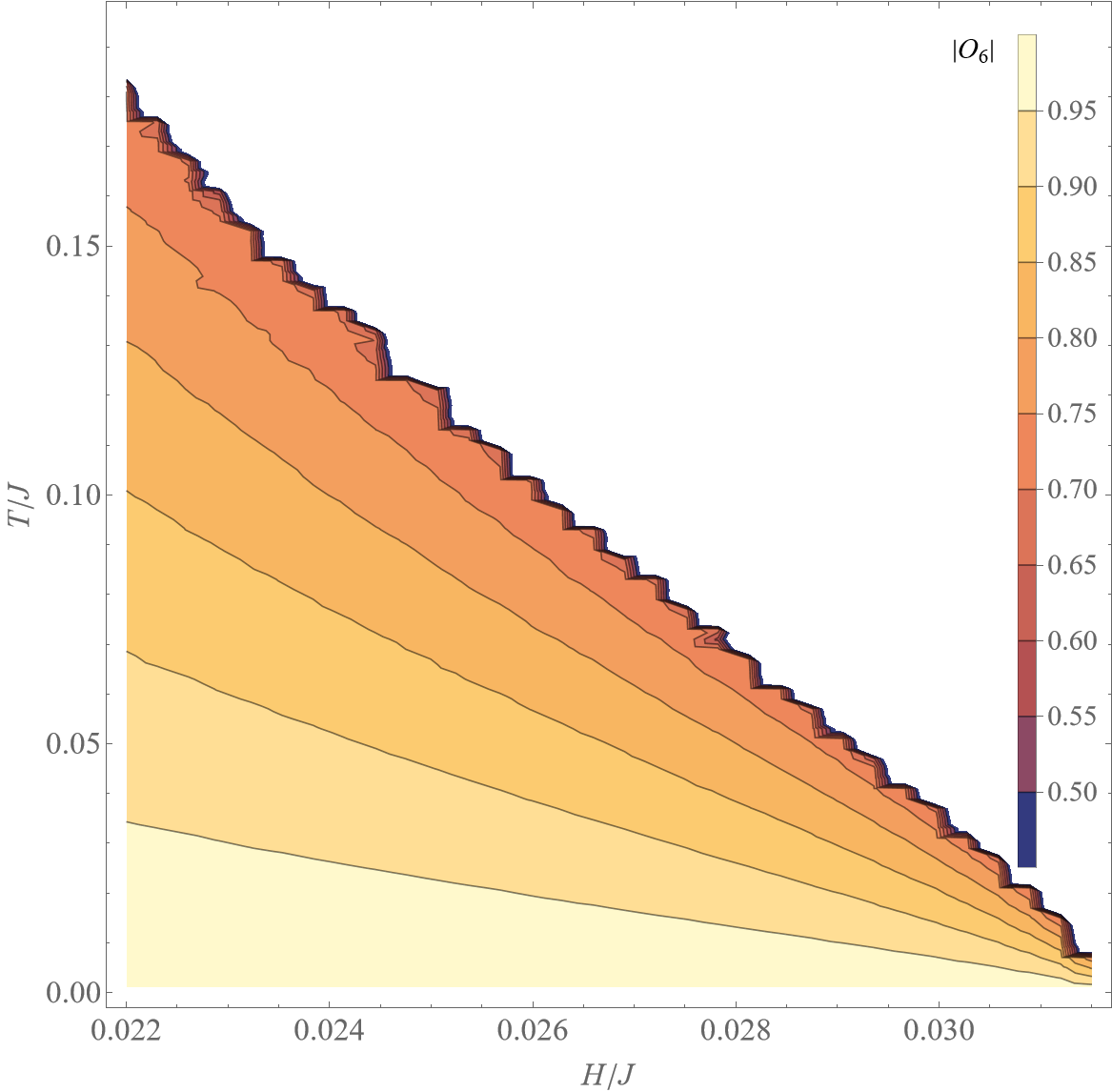}
\par\end{centering}
\caption{The contour plot showing the magnitude of the orientational order
parameter $O_{6}(H,T)$, assuming the concentration of skyrmions equal
to their equilibrium concentration at $T=0$ for any value of $H$. }

\label{Fig_PD_equilibrium_contour}
\end{figure}

Unlike the system of particles, the system of skyrmions can show the
melting transition not only on temperature $T$ but also on the magnetic
field $H$. The latter dependence comes both from the dependence of
the magnetic-field length $\delta H$, defined below Eq. (\ref{U_interaction}),
and from the dependence of the skyrmion's core energy $\Delta E(H)$
that can be computed from the lattice-spins model. Assuming the concentration
of skyrmions in the system to be equal to their equilibrium concentration
at $T=0$ found in Sec. \ref{Sec_interaction}, one can compute the
$\left(H,T\right)$ phase diagram by increasing the temperature from
zero for each value of $H$, performing the Monte Carlo simulation,
and detecting the melting temperature $T_{m}$. The results of such
computation for a small point-particle system with $N_{S}=1080$ and
pbc are shown in Fig. \ref{Fig_PD_equilibrium} together with the
skyrmion-skyrmion repulsion energy $U_{0}$. The same data is shown
in Fig. \ref{Fig_PD_equilibrium_contour} which gives the idea of
the values of $|O_{6}(H,T)|$. 

\begin{figure}
\begin{centering}
\includegraphics[width=8cm]{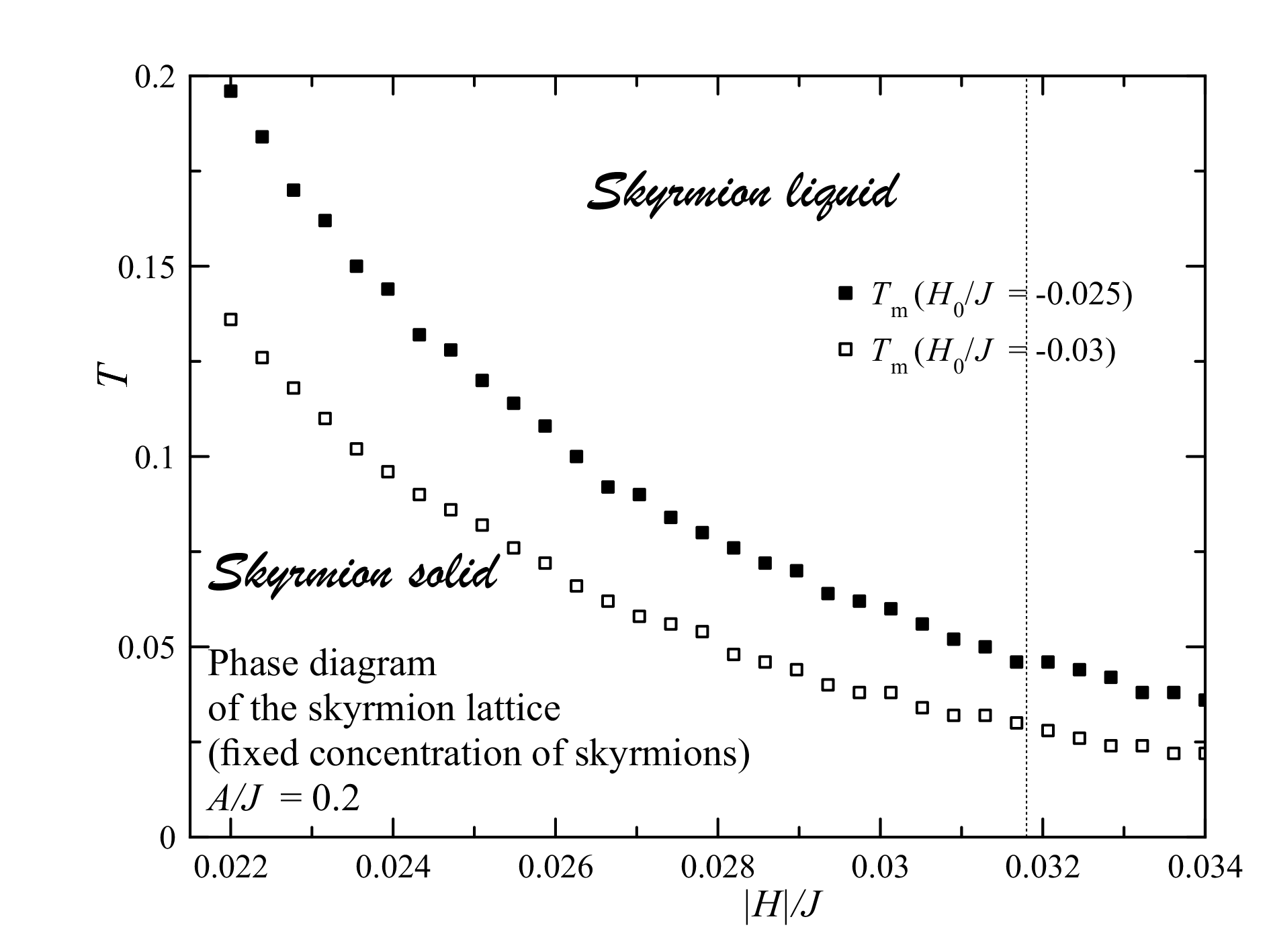}
\par\end{centering}
\caption{The phase diagram of the skyrmion liquid-solid system for the two
values of the concentration of skyrmions equal to their equilibrium
concentrations at $T=0$ for two different values of $H$. Here, for
other values of $H$, the concentration of skyrmions is non-equilibrium.}

\label{Fig_PD_non-equilibrium}
\end{figure}
As pointed out above, the number of skyrmions in the system is difficult
to change at low temperatures including the region of the melting
of the skyrmion lattice. With $N_{S}$ decoupled from the other parameters,
one has to consider the three-parameter phase diagram $\left(H,T,N_{S}\right)$.
Leaving this task for now, one can produce $\left(H,T\right)$ diagrams
with $N_{S}$ fixed. Fig. \ref{Fig_PD_non-equilibrium} shows the
latter for the two values of the skyrmion concentration corresponding
to the equilibrium at $T=0$ and the two different values of the magnetic
field $H_{0}$ computed via the minimization of the energy in Sec.
\ref{Sec_interaction}. Here, the skyrmion lattice exists also in
the region to the right of the vertical dotted line where the skyrmion's
core energy $\Delta E$ is positive and skyrmions are metastable.
At $|H|/J=0.034$ skyrmions collapse. 

\subsection{Skyrmion lattice in the gradient magnetic field}

\label{Sec_HGrad}

\begin{figure}
\begin{centering}
\includegraphics[width=8cm]{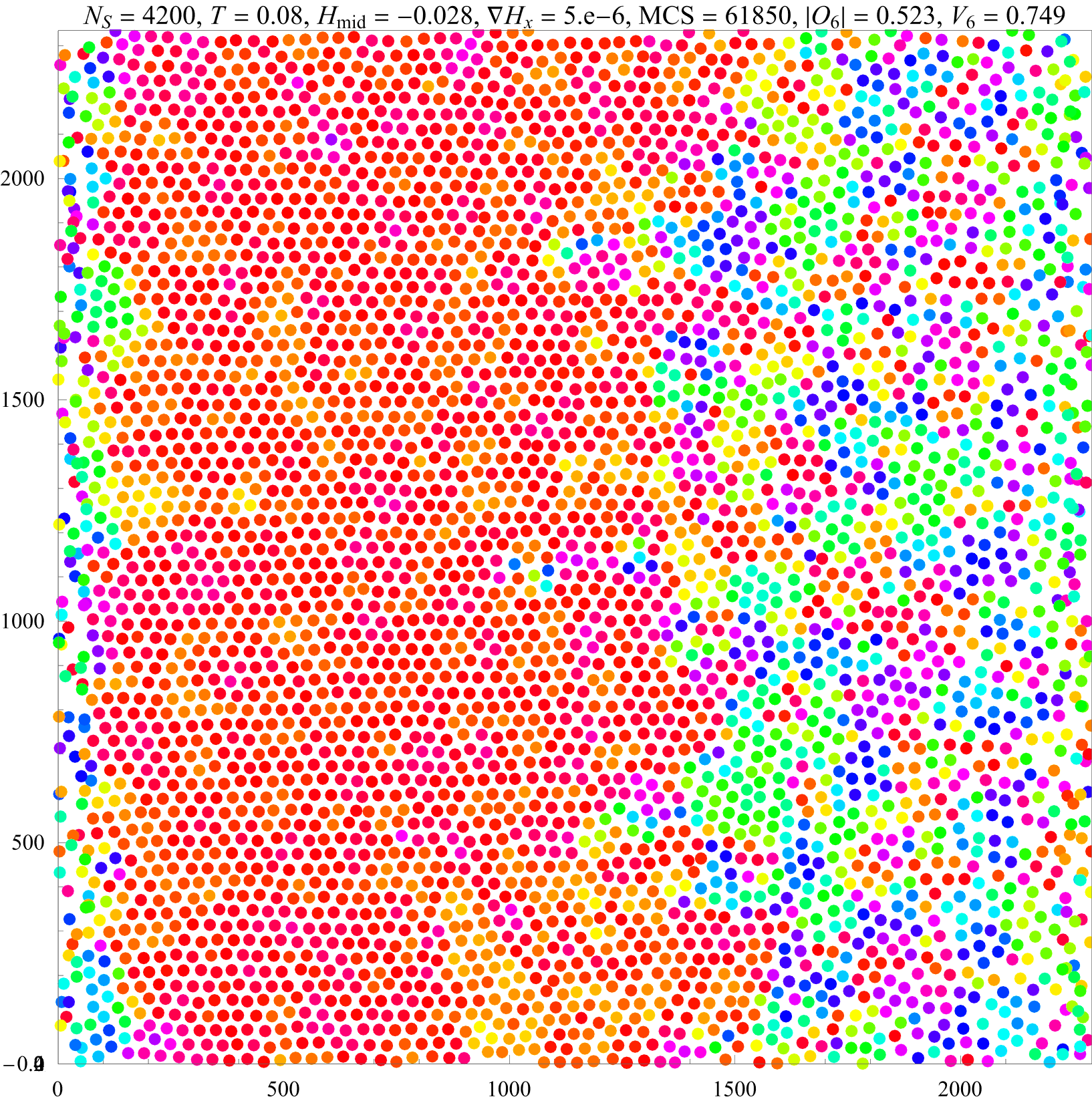}
\par\end{centering}
\caption{Skyrmion configuration with a solid-liquid interface in the presence
of a magnetic-field gradient. On the left, the magnetic field is weaker,
the skyrmion repulsion is stronger, $T_{m}$ is higher, and at the
given temperature the skyrmion lattice holds. On the right, the field
is weaker and the lattice melts.}

\label{Fig_Lattice_with_HGrad}
\end{figure}
\begin{figure}
\begin{centering}
\includegraphics[width=8cm]{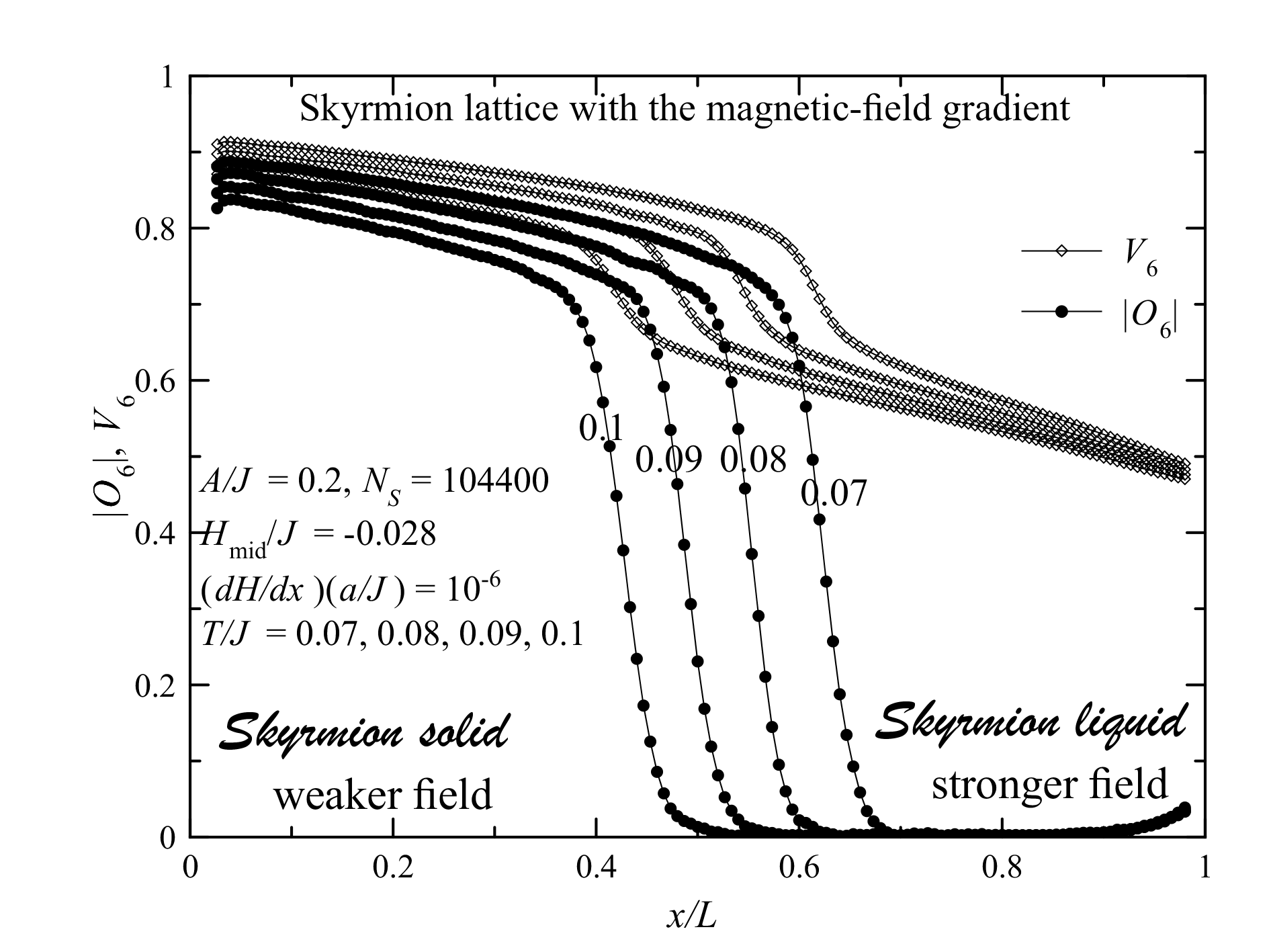}
\par\end{centering}
\caption{The magnitude of the orientational order parameter $O_{6}$ and the
hexagonality value $V_{6}$ in the presence of a magnetic-field gradient
at different temperatures show a solid-liquid interface.}

\label{Fig_O6V6_with_HGrad}
\end{figure}

The influence of the magnetic field on skyrmions is impressively demonstrated
by the solid-liquid interface caused by the gradient magnetic field,
see Fig. \ref{Fig_Lattice_with_HGrad}. For the temperatures in the
vicinity of $T_{m}$ applying a field gradient causes phase separation
because the melting temperature depends on the magnetic field. We
illustrated this phenomenon by applying the maximal possible field
gradient so that near the left boundary of the system the field is
only a bit stronger than the value at which skyrmions are branching
out and give way to the laminar domain structure and near the right
boundary the field is only a bit weaker than the skyrmion-collapse
field. For $A/J=0.2$ the field in the middle of the system is $H_{\mathrm{mid}}/J=-0.028$.
For the small system of $N_{S}=4200$ skyrmions shown in Fig. \ref{Fig_Lattice_with_HGrad}
the maximal value of the gradient we used was $\left(\partial H_{z}/dx\right)(a/J)=5\times10^{-6}$. 

Fig. \ref{Fig_O6V6_with_HGrad} shows quantitative results for the
skyrmion solid-liquid interfaces at different temperatures for a large
system of $N_{S}=104400$ skyrmions in the presence of the maximal
magnetic-field gradient $\left(\partial H_{z}/dx\right)(a/J)=10^{-6}$.
Here, the values of $|O_{6}|$ and $V_{6}$ are averages over the
bins with different $y$ positions, keeping the $x$ positions as
a parameter.

Both Figs. \ref{Fig_Lattice_with_HGrad} and \ref{Fig_O6V6_with_HGrad}
were obtained within the point-particle model that allows simulation
of large systems of skyrmions.

An interesting question is whether skyrmions would move to the right
or to the left if a gradient field is applied. It is known that a
gradient field can be used to move skyrmions. Within the PP model,
this effect comes from the magnetic-field dependence of the skyrmions
core energy, $\Delta E(H)$. As $\Delta E$ is lower for weaker fields,
skyrmions would move to this region, on the left in the figures. However,
skyrmions also repel each other, and in the region of the weaker field,
the repulsion is stronger as the magnetic length defining the skyrmion-skyrmion
interaction is longer. Under the action of the increased repulsion,
skyrmions would move away from the region of the weaker field. These
two competing effects compensate each other so that one cannot see
any non-uniform density of skyrmions in Fig. \ref{Fig_Lattice_with_HGrad}.
This is a plausible result because the equilibrium concentration of
skyrmions that was set in the initial state used to obtain the results
above is the result of the balancing between the skyrmion's core energy
and their repulsion energy. One can expect that for the concentration
of skyrmions lower than the equilibrium one, skyrmions will shift
to the left as their repulsion is weaker and the core-energy effect
dominates. On the other hand, for the concentration of skyrmions higher
than the equilibrium one the repulsion is stronger and makes skyrmions
shift to the right.

\section{Conclusions}

\label{Sec_Conclusion}

We have studied temperature and field dependence of the orientational
order in 2D skyrmion lattices developed in the background of over
one million lattice spins. Two complementary approaches have been
investigated by the Monte Carlo method, the one in which skyrmions
are treated as point particles with repulsive interaction derived
from a microscopic model of lattice spins, and the other in which
skyrmions are extended objects that emerge in a full microscopic spin
model. Excellent agreement between the two models has been found that
provides strong confidence in our results. 

The long-range ordering in a hexagonal skyrmion lattice is strongly
influenced by the shape of the boundary because it favors certain
orientations of hexagons and, thus, generates an orientational epitaxy.
This, in turn, produces a finite-size effect on the long-range orientational
order. The effect is especially strong for a rhomboid shape of the
boundary that favors uniform orientation of hexagons and thus a monocrystalline
order. On the contrary, the boundary of the square shape generates
frustration that forces a monocrystalline hexagonal lattice to break
into a polycrystal. 

The hexagonal skyrmion lattice is stable in a finite temperature and
magnetic field range. At a fixed field, its melting is clearly seen
in a step-like drop of the orientational order parameter at a certain
temperature that depends on the field. At that temperature, the magnetization
of the system exhibits a kink, which provides an independent tool
for detecting the solid-liquid phase transition. At a fixed temperature,
the transition can be achieved by increasing the field. We have drawn
the temperature-field phase diagrams for different concentrations
of skyrmions. 

The existence of the solid-liquid phase transition on both temperature
and the magnetic field in a 2D skyrmion lattice introduces a new element
into the studies of the problem of 2D melting that has been investigated
for 50 years. In particular, we observed an interesting behavior when
we applied a gradient of the magnetic field to such a system. In the
presence of the field gradient, it develops an interface between solid
and liquid phases. This may provide another independent tool for observing
phases of the skyrmion matter in experiments.

\section{Acknowledgments}

This work has been supported by the grant No. DE-FG02-93ER45487 funded
by the U.S. Department of Energy, Office of Science.

\end{document}